\documentclass[twoside,twocolumn]{article}
\usepackage[super,sort&compress,comma]{natbib} 
\usepackage[version=4]{mhchem}
\usepackage{times,mathptmx}
\usepackage{sectsty}
\usepackage{balance} 

\usepackage{graphicx} %eps figures can be used instead
\usepackage{lastpage}
\usepackage[format=plain,justification=justified,singlelinecheck=false,font=small,labelfont=bf,labelsep=space]{caption} 
\usepackage{fancyhdr}
\usepackage{romannum}
\usepackage{color}
\usepackage{mathtools}
\usepackage{booktabs}
\usepackage{array,dcolumn}

\usepackage{url}
\usepackage{hyperref}

\begin{document}

\twocolumn[
\begin{@twocolumnfalse}
\noindent\LARGE{\textbf{Influence of mineralization and injection flow rate on flow patterns in three-dimensional porous media}}
\vspace{0.6cm}

\noindent\large{\textbf{R. Moosavi \textit{$^{a}$}, 
A. Kumar \textit{$^{a}$}, 
A. De Wit \textit{$^{b}$}, 
and M. Schr\"oter \textit{$^{a,c}$}}}\vspace{0.5cm}

\date{\today}

\noindent \normalsize{}
\vspace{0.5cm}
 \end{@twocolumnfalse}
 ]
\footnotetext{\textit{$^{a}$~Max Planck Institute for Dynamics and Self-Organization (MPIDS), 37077  G\"ottingen, Germany}}
\footnotetext{\textit{$^{b}$~Universit\'e  libre de Bruxelles (ULB), Nonlinear Physical Chemistry Unit, CP231, 1050 Brussels, Belgium
}}
\footnotetext{\textit{$^{c}$~E-mail: matthias.schroeter@ds.mpg.de}}

\begin{abstract}
Reactive flows inside porous   media play an important role in a number of geophysical 
and industrial processes. Here we present  three-dimensional experimental measurements on how precipitation and flow patterns change with the flow rate inside a model porous medium consisting
of monodisperse glass beads.
The sample is initially filled with an aqueous solution of sodium carbonate
into which a solution of barium chloride is injected at a constant flow rate. Upon contact and reaction, the two reactants produce water-insoluble barium carbonate which precipitates onto the 
glass beads. This precipitate then modifies the flow morphology which in turn
changes the spatial distribution of the precipitate. We discuss the influence of 
the flow rate on the morphology of the flow pattern and demonstrate that neither viscous fingering  nor the 
Rayleigh-Taylor instability have any significant influence in our model system. 
\end{abstract}

%%%%%%%%%%%%%%%%%%%%%%%%%%%%%%%%%%%%%%%%%%%%%%%%%%%%%%%%%%%%%%%%%%%%%%%%%%%%%%
\section{Introduction}

In a context of global warming, there is increased need to explore the possibility of negative emissions technologies (NETS) aimed at decreasing the concentration of CO$_2$ in the atmosphere below the levels related to the simple stop of further release of the greenhouse gas in the air. Among these NETS, Carbon Capture and Sequestration (CCS) techniques attract increased attention as their objective is to capture CO$_2$ at the exit of industrial plants and inject it into soils where it should  ideally be safely trapped \cite{IPCC}. In this context, understanding in particular mineralization of CO$_2$ i.e. its reaction with components present in the soil to be transformed into solid phases is particularly relevant to explore the potential and security of sequestration techniques \cite{kaszuba:13,jun13,roc15}. Recent field experiments on the CARBFIX project in Iceland have shown that the injection of CO$_2$ mixed with water into basaltic rocks rich among others in calcium ions has induced a transformation of up to 95$\%$ of the injected CO$_2$ into minerals in less than 2 years \cite{matter:16}. The fact that CO$_2$ is first dissolved in water favors the formation of carbonate ions that are then readily available to react with the calcium ions according to the mineralisation reaction 
\begin{equation}
  \label{eq:precip_ca}
  \text{Ca}_{\text{(aq)}}^{2+} +  \text{CO}_{\text{3(aq)}}^{2-} \longrightarrow \text{CaCO}_{\text{3(s)}}
\end{equation}

The fact that the mixing occurs under injection conditions certainly also has an effect on the efficiency of the precipitation reaction. In this context, there is need to understand to what extent the spatio-temporal distribution of precipitate patterns in 3D porous media varies with the injection flow rate. 

Some model experiments have been conducted in confined Hele-Shaw geometries (two glass plates separated by a thin gap) \cite{tarta,werth,whi12,schu16,schu16b,brau17}. It has, among others, been  shown that the amount and spatial distribution of CaCO$_3$ precipitates varies with concentration and injection flow rates when an aqueous solution of carbonate is injected radially into an aqueous solution of Ca$^{2+}$ ions \cite{schu16,schu16b,brau17}. The same experiments conducted with calcium replaced by barium ions showed that in a large range of parameter values, the patterns are similar \cite{schu16c}. Hence, experiments with Barium, which is easier to follow by X-ray tomography, are representative in some limits of the dynamics with Calcium. 

While these experiments in confined Hele-Shaw cells already show that a simple precipitation reaction can be profoundly affected by the flow, the situation is even more complex in real 3D porous media. 
Three-dimensional porous media are in general opaque to visible light, the only  
exception being indexed matched model systems \cite{krummel:13}.
Consequentially, X-ray tomography 
has become a standard tool
\footnote{Other 3D imaging techniques used for the study of porous media are Magnetic Resonance Imaging (MRI)\cite{rose:13}  and neutron tomography \cite{murison:15}. MRI provides superior time resolution compared to to X-ray tomography, neutron tomography increased chemical sensitivity. However neither of these two methods reaches the spatial resolution of X-ray tomography.}
for all experiments where the interaction between different solutions inside a porous matrix is studied \cite{wildenschild:13}. 
Important examples, especially in the context of CCS techniques, are X-ray tomography studies on how carbonate rocks dissolve due to the injection of CO$_2$ saturated brine \cite{luquot:09,gouze:11,luhmann:14,luquot:14,vialle:14,garcia-rios:15,menke:16,qajar:16,al-khulaifi:17,lebedev:17,selvadurai:17,menke:18}. 
The same brine  with added H$_2$SO$_4$ not only dissolves carbonate rocks but also precipitates gypsum \cite{thaysen:17}.
Other examples include  the formation of salt crystals due to the evaporation of the salt containing brine \cite{desarnaud:15}, viscous fingering in porous media \cite{suekane:17},
diffusion driven reactions forming precipitates of gypsum %(CaSO$_4$)
and barite \cite{rajyaguru:19},  %(BaSO$_4$)
and  microbially induced carbonate precipitation \cite{minto:17}. 

However, only a small number of studies has leveraged the local resolution provided by tomography to study how precipitation reactions inside the porous medium can influence the flow itself.
Noiriel \textit{et al.} \cite{noiriel:16} and Godinho \& Withers \cite{godinho:18} studied the precipitation of calcite from supersaturated solutions injected into porous media and modelled the change in the flow field numerically. Cil \textit{et al.} compared the same type of setup with the calcite precipitation originating from two parallel streams of reactants inside the porous medium and found that
mixing patterns coevolve with the microstructure of precipitate \cite{cil:17}.

In this context, we study here experimentally using X-ray tomography how reactive flow patterns within a porous matrix vary when the injection flow rate is changed and a precipitation reaction forming insoluble barium carbonate takes place when a solution of sodium carbonate is displaced upwards by a solution of barium chloride. We find that, depending on the flow rate, the displacement takes place as a plug flow or along preferential channels. We discuss these findings in terms of a competition between precipitation that induce clogging of given pores and transport phenomena. 

%===========================
\section{Materials and Methods}
\subsection{Experimental approach}\label{setup}
Experiments are carried out in a cylindrical Plexiglass cell of $ 50~\text{mm} $ height and $16~\text{mm}$ inner diameter. The cell is filled with soda-lime glass spheres (MoSci) of diameter $ 125-150~\text{$\mu$m} $.
The spheres are confined between two hydrophilic membranes (Porex XS-8259, pore size 7-10 $\mu$m), 
the packing fraction is 0.59 $\pm 0.01$. Initially, the glass beads are
immersed in a 2.36 M aqueous solution of sodium carbonate ($ \text{Na}_{2}\text{CO}_{3} $).
Any  remaining air is removed by degassing the sample under vacuum. 

\begin{figure}[tbp]
\centering
\includegraphics[width=0.37\textwidth]{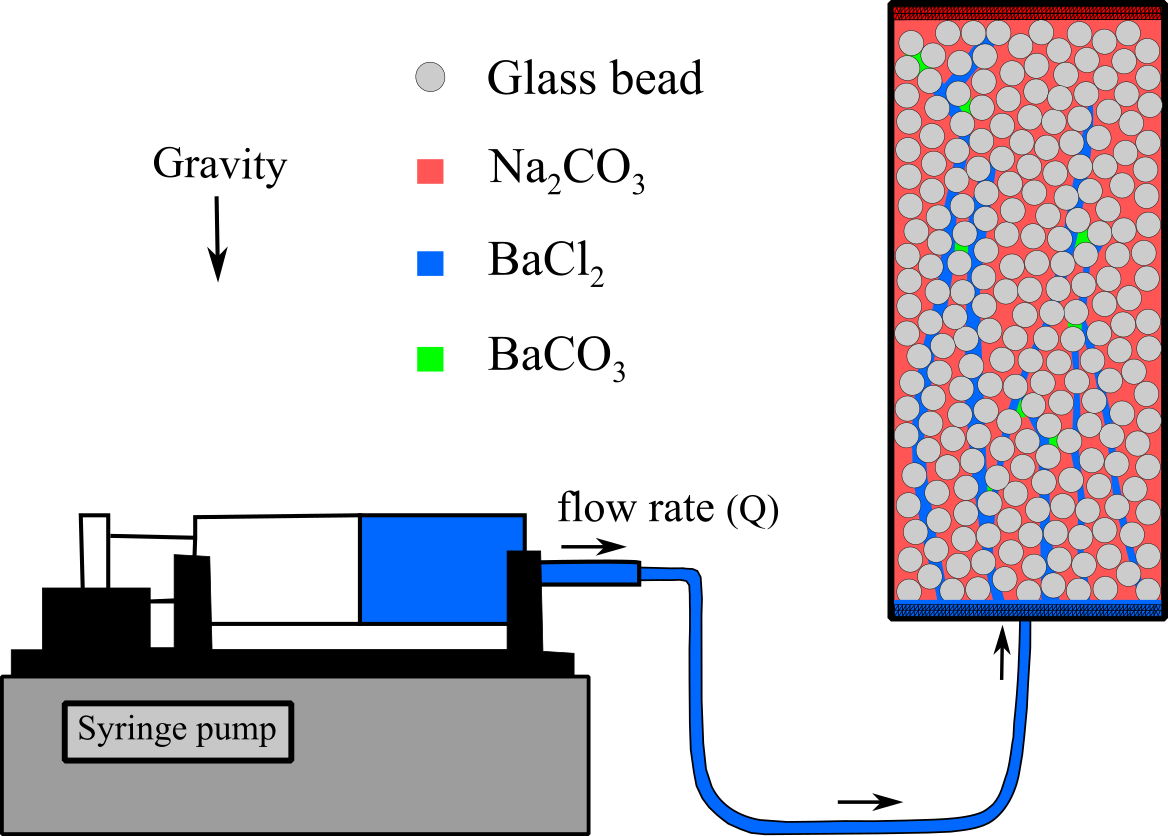}
\caption{Sketch of the experimental setup. A barium chloride solution is pumped into a
porous medium filled with a sodium carbonate solution. The reaction between these two solutions 
forms water-insoluble barium carbonate.}
\label{fig:setup}
\end{figure}

The experiment starts by injecting upwards a 1.78 M aqueous solution of Barium Chloride ($ \text{Ba}\text{Cl}_{2} $)
into the cell using a syringe pump (Harvard, PhD Ultra), cf.~figure \ref{fig:setup}. 
Due to the confining membrane, the injected fluid can enter the sample in an area corresponding to the full 
cross-section of the cylinder. Experiments are performed at different  flow rates $Q$ in the range of 0.01 - 0.6  ml/min, 
with three independent experiments for each value of $Q$. 
If the flow pattern inside the porous matrix is a
plug flow, these flow rates  translate into average front velocities 
$V$ of  2 - 124 $\mu$m/s. This range compares well to the front speed of 14  $\mu$m/s in a 
recent field scale experiment on carbon sequestration \cite{matter:16}.

As soon as the two reactive solutions get into contact, water-insoluble  $\text{Ba}\text{CO}_{3} $ precipitate is formed by the reaction:
\begin{equation}
  \label{eq:precip_ba}
  \text{Ba}_{\text{(aq)}}^{2+} +  \text{CO}_{\text{3(aq)}}^{2-} \longrightarrow \text{BaCO}_{\text{3(s)}}
\end{equation}
This precipitate then modifies the flow field which in turn changes the interface geometry between the 
two solutions. After an injection of 1.6 ml  of the $ \text{Ba}\text{Cl}_{2} $ solution, the flow is stopped and an 
X-ray tomography of the sample is taken for further analysis. 

The choice of $ \text{Ba}\text{Cl}_{2} $ as the reactant solution is motivated
by the  X-ray attenuation coefficient of barium ions which is 20 times higher than that of 
calcium ions (at 100 keV and equal molar concentration \cite{nist_db}).  
This facilitates the tomographic imaging of the flow patterns. As it has  been recently shown experimentally that precipitation patterns of barium carbonate are similar to those of calcium carbonate in Hele-Shaw cells at low flow rates such as those used here \cite{schu16c}, the present study is relevant to understanding CO$_2$ mineralization important in CCS techniques \cite{matter:16}.
The concentrations of the salt solutions are  chosen to ensure a good image contrast and a small difference
in density $\rho$. The physical properties of the two solutions are listed in table~\ref{tb:solution}.

\captionsetup[table]{aboveskip=0pt} 
% \captionsetup[table]{belowskip=-10pt}
\begin{table}[h!]
\centering
\caption{ Properties of the two reactant solutions. Kinematic viscosities $ \nu $ are measured 
with an Ubbelohde viscosimeter, densities $\rho$ using a glass pycnometer. }
\label{tb:solution}
\begin{tabular}{cccc}
  \midrule
  solution & $ \rho $ ($g/cm^{3} $) & conc. (M) & $ \nu $ ($ \text{mm}^{2}/\text{s} $) \\ 
  \midrule
  $\text{Na}_{2}\text{CO}_{3} $ & 1.2106 $\pm$ 0.0002 & 2.36 & 2.40 $\pm$ 0.02\\  %  2.398 $\pm$ 0.016  
  $ \text{Ba}\text{Cl}_{2} $ & 1.2691 $\pm$ 0.0008 & 1.72 & 0.89 $\pm$ 0.01\\    %0.888 $\pm$ 0.012
  \midrule
\end{tabular}
\end{table}

\subsection{X-ray tomography}
\label{x-ray}
In order to visualize the spatial distribution of the injected  $\text{BaCl}_{\text{2}}$ solution in three dimensions, we acquire X-ray tomographies (Nanotom, GE Sensing and Inspection) of each sample.  
X-rays are created using a tungsten target and an acceleration voltage of  140 kV.
Tomograms are reconstructed from $1620$ projections with 1152 $\times$ 1152  pixels each.
The  resulting three-dimensional volume is composed of voxels\footnote{The 3D equivalent of a pixel.}  
with a side length of 32 $\mu$m. This resolution allows us to visualize 36 mm of the total length of the 
sample directly above the filter.
The total acquisition time for one tomography is two hours.
We tested the stationarity of the final flow pattern by comparing 
X-ray tomographies of the same sample acquired directly after the 
experiment and two weeks later. No change in the morphology of the pattern was observed.

\begin{figure}[tbp]
\centering
\includegraphics[width=0.37\textwidth]{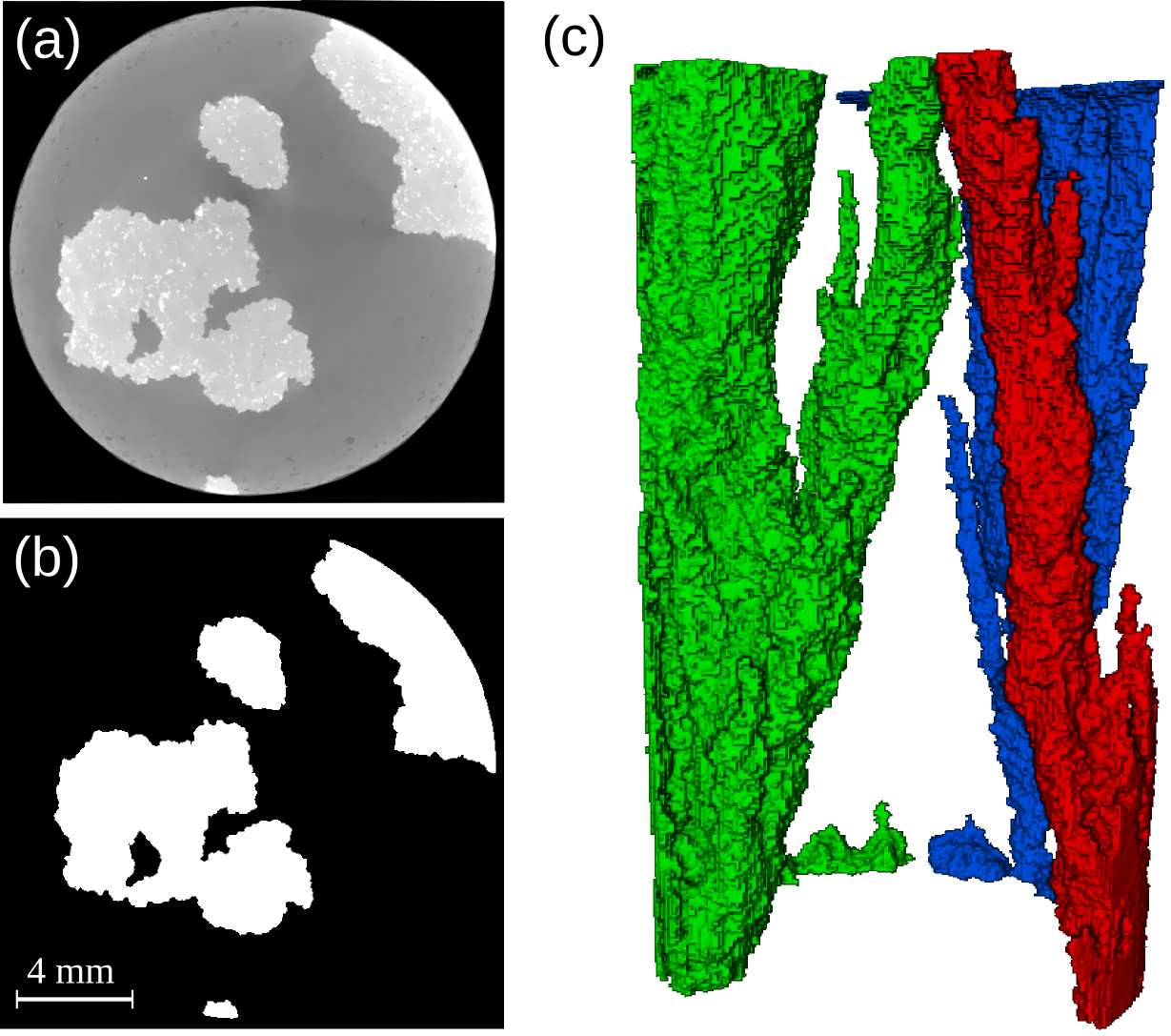}
\caption{Identification of the morphology of the injected $\text{BaCl}_{\text{2}}$ solution for an experiment at $Q=0.015 $ ml/min. 
a) Horizontal cross section through the  X-ray tomogram. Dark gray areas correspond to parts of the packing
containing  $ \text{Na}_{2}\text{CO}_{3} $ solution, bright gray areas contain the 
injected $\text{BaCl}_{\text{2}}$ solution. 
Inside the bright gray areas, precipitated $\text{BaCO}_{\text{3}}$ is visible as white pixels. Our 
resolution is not sufficient to analyze the precipitated $\text{BaCO}_{\text{3}}$ separately.
In the following, it is therefore combined with the $\text{BaCl}_{\text{2}}$ solution.
b) Binarization separates the area containing $\text{BaCl}_{\text{2}}$ solution (white) from the rest of the sample.
c) Using the full vertical stack of segmented images, the three-dimensional structure of the injected
 $\text{BaCl}_{\text{2}}$ solution can be visualized. The different colors distinguish  
separate fluid fingers.}
\label{patterns}
\end{figure}

Figure~\ref{patterns}a shows a horizontal two-dimensional slice of a tomography. The different gray values
correspond to glass spheres immersed in sodium carbonate solution (dark gray), glass spheres immersed in
barium chloride solution (bright gray), and barium carbonate which has precipitated onto the glass beads (white).

%%%%%%%%%%%%%%%%%%%%%%%%%%%%%%%%%%%%%%%%%%%%%%%%%%%%%%%%%%%%%%%%%%%%%%%%%%%%%%%%%%%%%%%%%%%

\subsection{Image analysis}\label{image-process}
The  segmentation of the 3D volume data into the two partial volumes filled by the
two liquid solutions is based on the gray values of the individual voxels. Two preprocessing steps 
facilitate \cite{weis:17}
this binarization: First, a bilateral filter (implemented in Avizo Fire) reduces the gray level noise while 
preserving the edges between areas. Second, multiplication with a radially varying factor (Matlab) removes  
the so called beam-hardening artifact which makes the interior of the cylinder appear less bright than its circumference.
Then the  3D volume is segmented with a threshold determined using Otsu's algorithm, effectively rendering all volume, 
which has been invaded by $\text{BaCl}_{\text{2}}$ solution, white (figure~\ref{patterns}b).
In a final step, isolated binarization artifacts smaller than 0.5 $\text{mm}^{3}$
are removed with a morphological transformations called opening (Avizo Fire).
A three-dimensional visualization of the  morphology of an injected  $\text{BaCl}_{\text{2}}$ solution
is shown in figure~\ref{patterns}c.

In order to characterize the flow morphology, we compute two measures for each tomography:
the number of discrete objects $ \text{N}_{\text{obj}} $  formed by the injected fluid and 
the surface to volume ratio $\sigma$ of these individual objects. To identify separate objects,
the set of all white voxels is split into subsets of voxels which are mutually connected by their faces (Avizo Fire).    
In figure~\ref{patterns}c, three such structures are identified by different colors.

Next we compute for each structure a) the total volume  $\text{V}_{3\text{D}} $ as the number of voxels belonging to it
and b) the  surface $\text{A}_{3\text{D}}$ enclosed by the structure  as the number of all its voxels which have at least 
one black voxel as neighbor. 
The so  obtained surface to volume ratio $ \sigma = \text{A}_{3\text{D}}/\text{V}_{3\text{D}} $
is dimensionless as both $\text{A}_{3\text{D}} $ and $\text{V}_{3\text{D}} $ are measured in voxels units. 
The final $\sigma$ value of a tomogram is obtained by averaging over all  $ \text{N}_{\text{obj}} $ structures contained in it.

%%%%%%%%%%%%%%%%%%%%%%%%%%%%%%%%%%%%%%%%%%%%%%%%%%%%%%%%%%%%%%%%%%%%%%%%%%%%%%%%%%%%%%%%%%%%%
\section{Results and discussion}\label{results}

\begin{figure*}[htbp]
\centering
\includegraphics[width=0.95\textwidth]{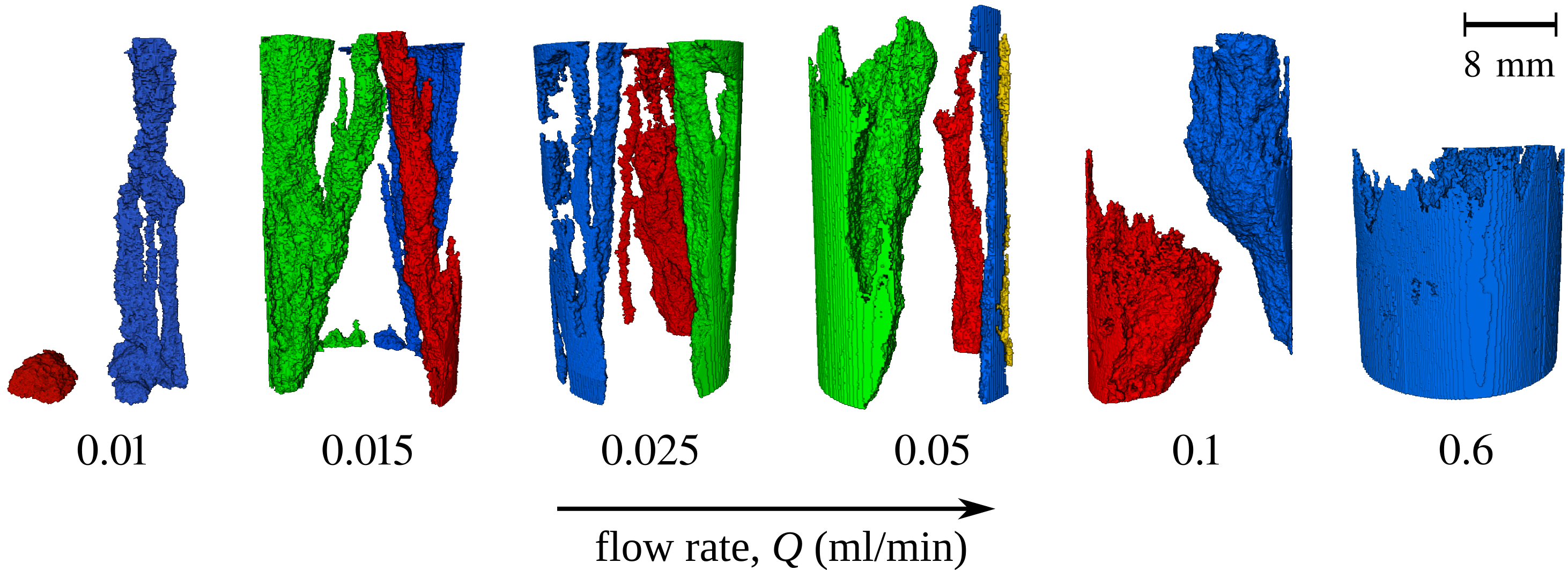}
\caption{The shape of the flow channels formed by the $\text{BaCl}_{\text{2}}$ solution   
depends on the flow rate $Q$: as $Q$ decreases, the structures become more ramified and fingered while a plug-like flow is observed for $Q>$0.1 ml/min. 
In all experiments, the  $\text{BaCl}_{\text{2}}$ solution is  injected from below into the solution of Na$_2$CO$_3$. Colors indicate individual separate flow channels.
}
\label{3d-patterns}
\end{figure*}
%----------------------------------------------
\subsection{The flow rate controls the flow pattern}
\label{sec:flow rate}
\begin{figure}[bp]
\centering
\includegraphics[width=0.40\textwidth]{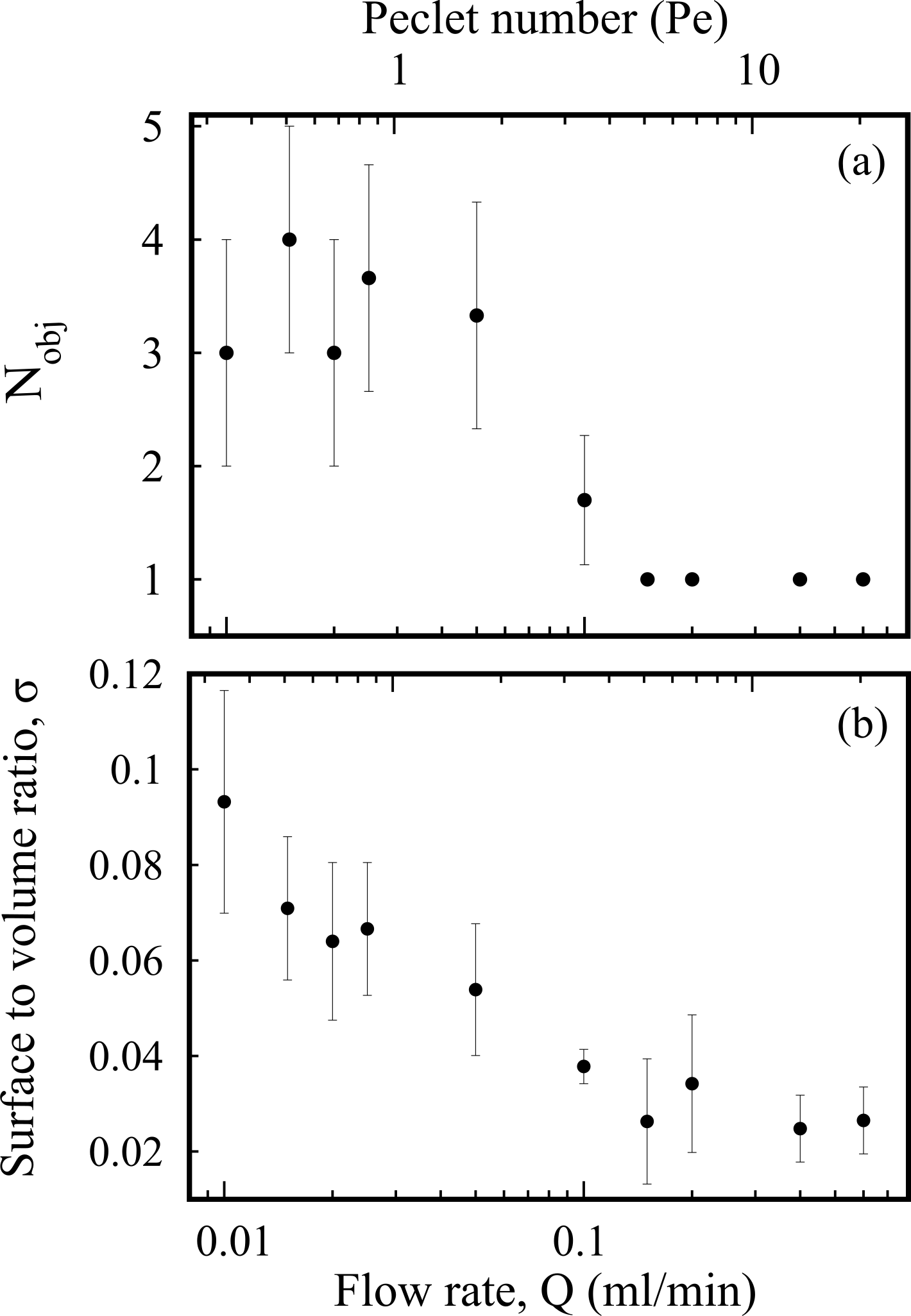}
\caption{Characterization of the transition to a plug-like flow above $Q \approx$ 0.1ml/min.
(a) Number $N_{\rm obj}$ of discrete streams formed by the injected  $\text{BaCl}_{\text{2}}$ solution 
(indicated by different colors in figure~\ref{3d-patterns}) and  
(b) surface to volume ratio $ \sigma$ as a function of flow rate. Larger values of 
$ \sigma$ indicate more ramified structures. Both  $ \sigma$
and  $N_{\rm obj}$ reach a lower plateau for  $Q >$ 0.1ml/min.
All data points are averages over three experiments.
}
\label{A3toV3-NumofObj}
\end{figure}

\begin{figure*}[h]
\centering
\includegraphics[width=0.9\textwidth]{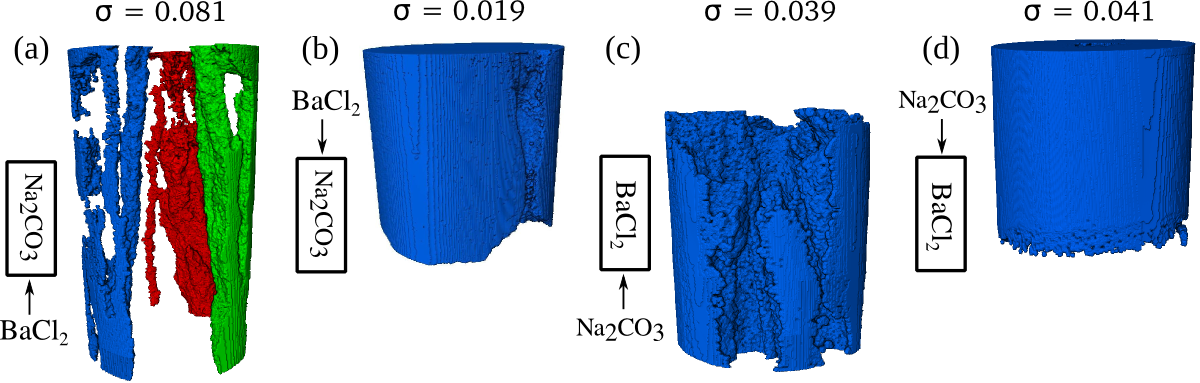}
\caption{Comparison of patterns obtained at $ Q = 0.025 ~\text{ml/min}$ depending on the direction of the flow in the gravity field and which solution is the displacing one. The surface to volume ratio values are written on the top of the images. The most rough and non-uniform structure is obtained when the less viscous and denser   solution of $\text{Ba}\text{Cl}_{2} $ is injected upwards into the solution of Na$_2$CO$_3$, which is a hydrodynamically stable displacement (panel a). The other cases feature more homogeneous displacements which rules out the potential influence of viscous fingering or of the Rayleigh-Taylor instability in the dynamics. }
\label{reverse flow}
\end{figure*}

The main result of our tomographic analysis shown in figure~\ref{3d-patterns}
is that the flow pattern of the  $\text{BaCl}_{\text{2}}$ solution replacing
the $ \text{Na}_{2}\text{CO}_{3} $ solution changes with the flow rate $Q$.
For  values of $Q$ larger than 0.1 ml/min, the  $\text{BaCl}_{\text{2}}$ solution
advances as  a cylindrical, plug-like front.
For smaller $Q$ we observe a transition to more ramified flows, i.e. 
the  $\text{BaCl}_{\text{2}}$ solution moves preferentially 
through some channel-like structures and in this way bypasses large areas which remain filled with stagnant $ \text{Na}_{2}\text{CO}_{3} $ solution. 

A more quantitative characterization of the transition is given in Figure \ref{A3toV3-NumofObj} displaying
the number of discrete structures $N_{\rm obj}$ as a function of the flow rate. 
A plug like flow corresponds to  $N_{\rm obj} = 1$, larger numbers indicate the formation of multiple 
flow fingers. The transition between these two regimes occurs around $Q \approx$ 0.1 ml/min.
In figure \ref{A3toV3-NumofObj}b, the surface to volume ratio $ \sigma$ is shown as a function of $Q$.
Higher values of $\sigma$ indicate a higher degree of ramification of the flow pattern. In our experiments, $\sigma$ decreases monotonically when the flow rate increases, until it  reaches a plateau around  $Q \approx$ 0.1 ml/min, indicating the  transition to a single plug-like front.

The transition between these two flow patterns can be explained, as seen below, on the basis of a microscopic picture showing how the precipitate formed by the reaction changes the pore geometry. Before doing so, let us first confirm that hydrodynamic instabilities are not involved in the process of ramification.

%-------------------------------------------
\subsection{Influence of the flow direction}
\label{instability}

To confirm that hydrodynamic instabilities are not at the origin of the flow patterns observed, we have analyzed the possible influence of  viscous fingering (VF) or Rayleigh-Taylor (RT)  instabilities that could occur due to viscous and/or density contrasts between the solutions (Table~\ref{tb:solution}). The experiments are performed by injection of one of the two reactant solutions into 
glass beads immersed in the other solution either from the top or from the bottom of a vertical cell (Fig.\ref{reverse flow}). 

The VF instability arises when a less viscous fluid is displacing a more viscous one and induces a deformation of the interface into fingering patterns \citep{saffman:58,homsy:87}. The RT instability on the other hand is a buoyancy-driven instability arising when a denser fluid lies above a less dense one in the gravity field \cite{sharp:84}. It also induces fingering patterns that have been well studied in the framework of CO$_2$ convective dissolution \cite{tho17}. The most unstable case occurs when both instabilities are at play, which should be here the case when the less viscous but denser solution of BaCl$_2$ (see Table I) displaces downwards the more viscous, less dense solution of Na$_2$CO$_3$. As seen on Fig.\ref{reverse flow}b, this situation  yields however a stable downward moving plug-flow type of displacement. This pattern is actually similar to the one obtained when the more viscous but less dense solution of Na$_2$CO$_3$ is displacing the barium chloride solution upwards (Fig.\ref{reverse flow}c), which a potentially buoyantly unstable but viscously stable displacement. Similarly, the potentially viscously  unstable but buoyantly stable case  shown in Fig.\ref{reverse flow}d features a plug-flow like pattern. Only the case that is stable towards both instabilities (Fig.\ref{reverse flow}a) features a much more ramified flow structure. This suggests that the VF or RT are not at the origin of the filament precipitation pathways but that these are formed when the hydrodynamically stable displacement at low velocities allow for the reaction to be most effective. 

%--------------------------------------------------------
\subsection{Clogging as the microscopic origin of the flow pattern }
The precipitation of solid barium carbonate 
as described by equation (\ref{eq:precip_ba}) is a fast reaction \cite{schu16c}. 
Hence,  the rate of formation of new $\text{BaCO}_{\text{3}}$(s) is controlled by the 
transport processes by which new educts get in contact within the reaction zone at the miscible interface between the two reactant solutions. 
At very low flow rates, diffusion is expected to be the dominant transport process while hydrodynamic dispersion takes over when $Q$ increases. 
The relative contribution of these two processes can be characterized 
by the P\'eclet number $Pe$ \cite{vialle:14,pra15}:
\begin{equation}
    Pe = \frac{d V}{D}
    \label{eq:peclet}
\end{equation}
where $d$ is the average grain diameter, and $D = 0.8 \times 10^{-9}$ m$^2$/s is the diffusion coefficient of Barium ions in water \cite{rard:80}.  Our range of average front velocities $V$ therefore covers a range of P\'eclet numbers from 0.34 to 20.4 with the transition between the two morphologies occuring around $Pe$ =3.4.
The absolute values of these  P\'eclet numbers are subject to a systematic error because the average front velocity is only a proxy of the actual liquid velocities in the ramified flow field. Nevertheless they capture 
the change in the flow pattern in a qualitative way.

At low $Pe$, the time scale of the global advective transport is longer than the time scale of the local diffusive transport between the two reactant solutions. Hence, at small values of $Pe$ or equivalently small values of $Q$ and thus $V$,  the local formation of precipitate by diffusion and reaction has  
sufficient time to close the gaps between individual glass spheres.  This leads to the formation of "tubes" which 
then guide the advancing flow front. On the other hand, if the   $\text{BaCl}_{\text{2}}$ solution
is injected fast enough, $Pe$ is larger, the advective transport takes over and 
the  produced solid $\text{BaCO}_{\text{3}}$ is spread out over a larger volume. 
In consequence, the local volume of the precipitate  is not sufficient to seal off throats between glass spheres and form channels that way.
The injected solution spreads then homogeneously in all directions and a plug-like displacement is observed. These trends are coherent with fingering-like instabilities due to precipitation \cite{nag14} and with the fact that the effect of precipitation reactions on viscous fingering in Hele-Shaw cells has been shown to decrease when $Pe$ increases \cite{nag08}. 

\begin{figure*}[h]
\centering
\includegraphics[width=0.99\textwidth]{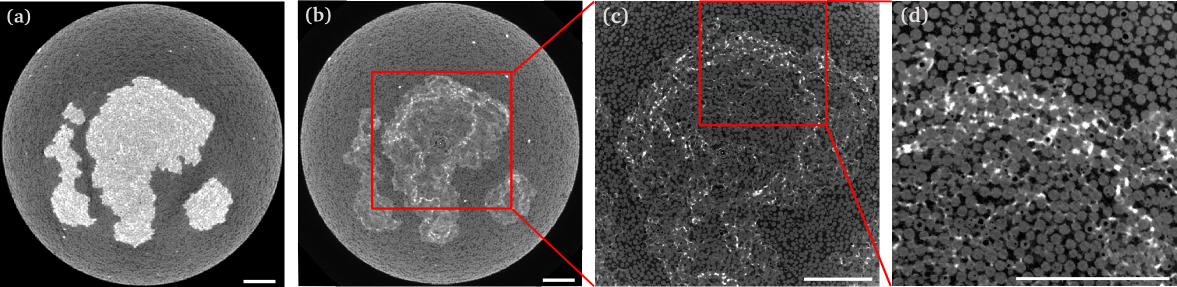}
\caption{Characterization of the barium carbonate deposit.
(a) Cross section of an X-ray tomography  of a sample before (a) and after (b)  washing out the
Barium chloride solution with water. Both image have a resolution of 32 $\mu$m/voxel. 
(c) Magnified region of interest taken from a new tomography at a resolution of 11 $\mu$m/voxel. 
(d) Optical zoom into a region of panel (c). Both panels (c) and (d) show that the
Barium carbonate is deposited in layers inside and at the circumference of the barium chloride 
flow finger. All scale bars are 2 mm long. 
}
\label{washing}
\end{figure*}

In order to study the viability of this hypothesis, we took additional tomographies with the aim of 
selectively visualizing the deposited  $ \text{Ba}\text{CO}_{3} $. Figure \ref{washing}a shows a sample into which 
a $ \text{Ba}\text{Cl}_{2} $ solution has first been injected at 
a flow rate of $ \text{Q} = 0.025~\text{ml/min} $ and has reacted with the sodium carbonate initially present. 
Then the system has been flooded with a volume of  pure water corresponding to 15 times the pore volume of the cell at a flow rate of $ \text{Q} = 0.05~\text{ml/min} $. This procedure removed all the remaining  $ \text{Ba}\text{Cl}_{2} $ solution. In consequence all white areas in figure \ref{washing}b,c,d represent solid $ \text{Ba}\text{CO}_{3} $.

A comparison of the X-ray tomographies before and after washing (figures~\ref{washing}a,b) shows
that the $ \text{Ba}\text{CO}_{3} $ deposit is organized in a layer like structure.  
The higher resolution tomographies shown in figures \ref{washing}c,d show that the deposit indeed closes 
off the space between individual glass spheres, in that way creating effective flow boundaries.
The fact that those layers are not only at the outer boundaries of the barium chloride flow finger visible
in figure \ref{washing}a is indicative of a stage-wise formation history of the finger: earlier boundaries
"break" with increasing pressure inside the finger. The spilled liquid then forms new boundaries at
geometrically favorable locations, creating an onion-like appearance of the cross-section.

\begin{figure}[h]
\centering
\includegraphics[width=0.45\textwidth]{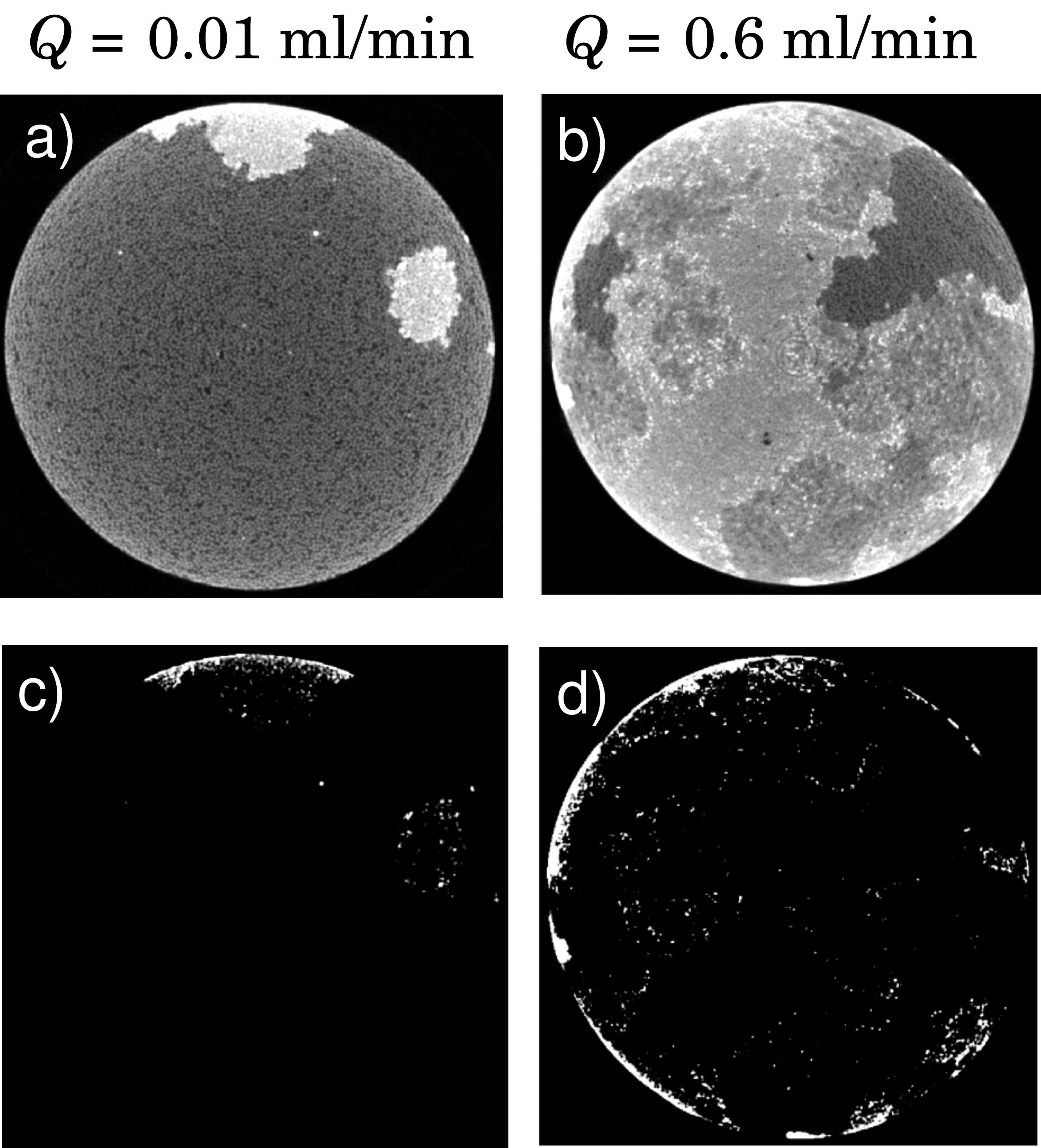}
\caption{The spatial distribution of the barium carbonate precipitate depends on the flowrate. Panels a) and b) show
cross sections ot the tomographies taken at the higest and lowest flow rate $Q$ applied in our experiment. Panels c) and d) have been binarized with a high threshold such that
only the barium carbonate precipitate remains visible.
}
\label{fig:bina_Q}
\end{figure}

Figure \ref{fig:bina_Q} confirms that the transition from the ramified to the  plug-like flow is also accompanied by a change of precipitation patterns.
In the top row of the image, cross-sections of tomographies  taken at the lowest and highest value of $Q$ are shown. If we make the reasonable assumption that the barium carbonate precipitate is the most dense and therefore most white part of the image, we can binarize these cross-sections with a rather high (but unfortunately arbitrary) threshold which results in the bottom row of figure \ref{fig:bina_Q}. Here the white area now represents the barium carbonate precipitate. At the high flow rate, the precipitate formed is distributed all over the full cross-section of the sample and therefore providing only very little guidance for the fluid flow, in agreement with the observed plug flow. At the low flow rate, the precipitate is concentrated in the finger-like flow structure.

%%%%%%%%%%%%%%%%%%%%%%%%%%%%%%%%%%%%%%%%%%%%%%%%%%%%%%%%%%%%%%%%%%%%%%%%%%%%%%%%%%%%%%%%%%
\section{Conclusions}
Three-dimensional reactive flow patterns have been analyzed here experimentally in porous media consisting in beads initially immersed in a solution of sodium carbonate. Upon injecting from bottom to top a solution of barium chloride, a precipitation reaction yields solid barium carbonate. It is observed by X-ray tomography that, at low injection flow rate, the flow pattern features ramified and irregular flow paths while at higher flow rates, a plug-flow displacement is obtained. A comparison of various displacements from either top to bottom or the reverse and alternating the displacing and displaced solutions shows that hydrodynamic instabilities such as viscous fingering or the buoyancy-driven Rayleigh-Taylor instability are here ruled out in the dynamics. Instead, a higher resolution tomography analysis suggests that, at low flow rate, diffusion is more effective than advection such that the precipitation reaction is more effective to produce the solid barium carbonate that, locally, blocks the pores. This, in turn, favors the growth of specific channeled pathways. At larger flow rates, advection dominates and a plug-like flow is observed as the precipitation is less efficient in blocking the pores. 
These  results confirm experimentally in 3D porous matrices the fact that the efficiency of mineralization and the spatio-temporal distribution of the solid phase are strongly affected by the injection flow rate. This paves the way to future work aiming at  optimizing CO$_2$ mineralization in flow conditions in real field conditions.

%%%%%%%%%%%%%%%%%%%%%%%%%%%%%%%%%%%%%%%%%%%%%%%%%%%%%%%%%%%%%%%%%%%%%%%%%%%%%%%%%%%%%%%%%%%%%%%%%%% 
\section{Acknowledgments}
We thank Markus Benderoth for experimental support.
R.M.~and M.S.~acknowledge financial support
from BP plc.~within the ExploRe program.
A.D.~acknowledges financial support from FRS-FNRS under the PDR CONTROL programme. 

\footnotesize{
\providecommand*{\mcitethebibliography}{\thebibliography}
\csname @ifundefined\endcsname{endmcitethebibliography}
{\let\endmcitethebibliography\endthebibliography}{}

}

\end{document}